\documentclass[journal]{IEEEtran}
\IEEEoverridecommandlockouts
\usepackage{lipsum} 
\usepackage[tight,footnotesize]{subfigure}
\usepackage{enumitem} 
\usepackage{float}
\usepackage{multicol,multienum}
\usepackage{breqn}
\usepackage{stfloats}
\usepackage{multirow}
\usepackage{diagbox}
\usepackage{hyperref}
\usepackage{bm}
\usepackage{cite}
\usepackage{graphicx}
\usepackage{tabularx}
\usepackage[ruled,vlined]{algorithm2e} 
\usepackage{booktabs}

\usepackage[tight,footnotesize]{subfigure}
\usepackage{url}
\usepackage{doi}
\usepackage{amssymb}
\usepackage{xcolor}
\ifCLASSINFOpdf

\else

\fi

\begin{document}
\title{Federated Learning in the Sky: Aerial-Ground Air Quality Sensing Framework with UAV Swarms}

\author{Yi~Liu,
	Jiangtian~Nie,
	Xuandi~Li,
	Syed Hassan Ahmed,
	Wei Yang Bryan Lim,
	and~ Chunyan Miao
	\thanks{Y. Liu is with the School of Data Science and Technology, Heilongjiang University, Harbin, China (email: 97liuyi@ieee.org). J. Nie, X. Li are with the School of Computer Science and Engineering, Nanyang Technological University, Singapore (emails: jnie001@e.ntu.edu.sg, cindy.li@ntu.edu.sg). S.H. Ahmed was with the Department of Computer Science, Georgia Southern University at the time of preparation of this work. (E-mail: sh.ahmed@ieee.org). W.Y.B Lim is with Alibaba Group and Alibaba-NTU Joint Research Institute, Nanyang Technological University, Singapore (email: limw0201@e.ntu.edu.sg). C. Miao is with School of Computer Science and Engineering, Nanyang Technological University, Singapore, and Joint NTU-UBC Research Centre of Excellence in Active Living for the Elderly (LILY) (email: cymiao@ntu.edu.sg). J. Nie is the corresponding author.}
}

\markboth{IEEE Internet of Things Journal}%
{}

\maketitle

\begin{abstract}
	Due to air quality significantly affects human health, it is becoming increasingly important to accurately and timely predict the Air Quality Index (AQI). To this end, this paper proposes a new federated learning-based aerial-ground air quality sensing framework for fine-grained 3D air quality monitoring and forecasting. Specifically, in the air, this framework leverages a light-weight Dense-MobileNet model to achieve energy-efficient end-to-end learning from haze features of haze images taken by Unmanned Aerial Vehicles (UAVs) for predicting AQI scale distribution. {Furthermore, the Federated Learning Framework not only allows various organizations or institutions to collaboratively learn a well-trained global model to monitor AQI without compromising privacy, but also expands the scope of UAV swarms monitoring.} For ground sensing systems, we propose a Graph Convolutional neural network-based Long Short-Term Memory (GC-LSTM) model to achieve accurate, real-time and future AQI inference. The GC-LSTM model utilizes the topological structure of the ground monitoring station to capture the spatio-temporal correlation of historical observation data, which helps the aerial-ground sensing system to achieve accurate AQI inference. {Through extensive case studies on a real-world dataset, numerical results show that the proposed framework can achieve accurate and energy-efficient AQI sensing without compromising the privacy of raw data.}
\end{abstract} 


\IEEEpeerreviewmaketitle

\section{Introduction}
According to a report from the World Health Organization, about 7 million premature deaths in 2012 were related to air pollution that increases the probability of serious diseases, e.g., respiratory and cardiovascular diseases for humans \cite{ref-1}. To quantify the severity degree of air pollution, a metric named Air Quality Index (AQI) is calculated by the concentration of various particulate matter (e.g., $\mathrm{PM_{2.5}}$ and $\mathrm{PM_{10}}$) in the air \cite{ref-3}. With the help of AQI, humans can take protective measures in advance. Therefore, it is significantly important for humans to achieve accurate and timely  AQI monitoring in order to  find efficient ways for air pollution control.

The existing AQI monitoring approaches are mainly divided into two categories: \textit{sensor-based monitoring} and \textit{vision-based monitoring} \cite{ref-2}. The \textit{sensor-based} monitoring approaches typically utilize sensor monitoring stations set up by public or private agencies on dedicated sites in a city to achieve AQI monitoring. Nevertheless, these approaches can only provide coarse-grained two-dimensional (2D) monitoring due to the long distance between two monitoring stations. To achieve fine-grained AQI monitoring, large-scale Internet of Things devices are applied to stationary stations in cities \cite{ref-27}. Although densely deployed static sensors can achieve high-precision monitoring, this way still suffers from  problems of high-costs and lack of mobility. Previous research has shown the potential that mobile devices (e.g.,  mobile phones, vehicles, and balloons) can access sensor data to monitor AQI \cite{ref-4,ref-6,ref-8}. However, mobile devices need to acquire a certain amount of air quality data for AQI monitoring using sensor-based approaches, which results in a large amount of energy consumption for traveling large-scale monitoring areas.


Unlike the sensor-based monitoring approaches, \textit{vision-based} monitoring methods include static station monitoring and mobile crowdsourcing \cite{ref-52}. Static monitoring stations infer AQI at restricted sites over the whole region by taking aerial images \cite{ref-9}. {Liu \textit{et al.} in \cite{ref-13} developed an AQI monitoring mobile application, which achieves fine-grained $\mathrm{PM_{2.5}}$ monitoring in a crowdsensing way.} Although the crowdsourcing-based monitoring method expands the scope of AQI monitoring, {its performance is usually affected by low-quality data sources and restricted by data privacy protection policies and schemes \cite{ref-12,ref-14,ref-57}.}
Previous researches have integrated sensor-based with vision-based methods to achieve accurate AQI monitoring, but have the following limitations:
\begin{enumerate}[label=(\roman*)]
	\item \textbf{High Energy Consumption}: For the sensor-based method, to achieve high-precision AQI monitoring, a large number of sensor devices are densely deployed on static stations in the city, which will cause expensive deployment costs and energy consumption. {Even if we use a small number of mobile devices for AQI monitoring, it still causes high energy consumption during moving because the devices need to obtain air quality data in a large scale areas \cite{ref-15,ref-53,ref-54}.} 
	\item \textbf{Small-scale Monitoring}: The monitoring range of the AQI monitoring station established by the government is less than 5\% of the urban area, which is not enough for the public to obtain real-time air quality inference \cite{ref-16}. {This generally requires many institutions to establish an AQI monitoring model with large monitor scope by sharing data \cite{ref-21}.}
	\item \textbf{Privacy Concerns}: Public and private agencies build AQI monitoring models collaboratively by sharing data obtained through crowdsourcing. However, the General Data Protection Regulation (GDPR) \cite{ref-17} prohibits direct sharing of user data between agencies due to privacy concerns, thus leading to data island problems. {Therefore, we need to achieve accurate AQI monitoring while protecting privacy.}
\end{enumerate}
These limitations hinder the widespread deployment and application of AQI monitoring systems in smart cities. {Therefore, it is vital to design an efficient framework with the properties of energy efficiency, better scalability, and privacy preservation for monitoring and inferring AQI values.}

{To address the above limitations, we propose a light-weight federated learning-based aerial-ground air quality sensing framework with Unmanned Aerial Vehicles (UAVs) swarms, to monitor and forecast AQI distributions {from} spatial-temporal perspectives \cite{ref-49,ref-50,ref-51}.} Specifically, we combine sensor-based with vision-based methods to achieve fine-grained 3D AQI scale inference with low energy consumption by using UAV. Unlike existing frameworks \cite{ref-55,ref-56}, we implement: (1) {mobile vision-based aerial sensing empowered by federated learning over UAV swarms, which infers the region-level AQI scale by utilizing a federated learning framework that deploys deep Convolutional Neural Networks (CNN) without sharing raw data;} (2) ground sensing over a wireless sensor network (WSN) for small-scale accurate spatial-temporal AQI inference, using a Graph Convolutional Neural Network-based Long Short-Term Memory (GC-LSTM) model; (3) an improved MobileNet \cite{ref-19} model, called Dense-MobileNet, is deployed in UAV that achieves high inference accuracy while significantly reducing energy consumption. 

The contributions of this paper are summarized as follows:
\begin{itemize}
	\item To address the privacy concerns issues in AQI monitoring, we build a federated learning (FL) framework on UAV swarms, which enables different agencies to collaboratively monitor AQI without sharing raw data.
	\item To achieve a fine-grained AQI scale distribution with low-energy consumption, we introduce a light-weight Dense-MobileNet model in FL framework, which can achieve energy-efficient end-to-end learning from haze features of haze images taken by UAVs to predict AQI scale distribution.
	\item The proposed GC-LSTM model utilizes the topology of the ground WSN system not only to accurately predict the AQI spatio-temporal value, but also to improve the performance of UAV vision-guided based method.
	\item We conduct extensive experiments on a real-world dataset to demonstrate the performance of the proposed schemes for AQI monitoring compared to non-federated learning methods.
\end{itemize}
The rest of this paper is organized as follows. Section \ref{sec-2} reviews the literature on AQI monitoring and federated learning research in UAV. Section \ref{sec-3} presents the overview of light-weight federated learning-based UAV vision-guided aerial-ground air quality sensing framework. Section \ref{sec-4} presents Dense-MobileNet model. In Section \ref{sec-5}, we presents GC-LSTM based AQI inference model. Section \ref{sec-6} discusse the experimental results. Concluding remarks are described in Section \ref{sec-7}.

\section{Related Work}\label{sec-2}
\subsection{AQI Monitoring Methods}
AQI monitoring has always been a hot issue in urban computing, which serves as a function of improving air quality and urban planning. Although researchers have proposed many new methods, they can generally be divided into two categories: \textit{sensor-based monitoring} and \textit{vision-based monitoring} approaches.

\subsubsection{\textbf{Sensor-based Monitoring Methods}} Government agencies set up a small number of stationary stations to infer AQI by collecting sensor data. This method is limited by the long distance between the stations and cannot achieve high-precision AQI inference. Although Aircloud \cite{ref-5} and U-air \cite{ref-6} achieve high-precision AQI monitoring by deploying high-density Internet of Things devices, their deployment carries high costs. To solve this problem, researchers use low-cost mobile sensing devices such as mobile phones and balloons for AQI monitoring \cite{ref-13,ref-15,ref-8}. However, none of them can achieve fine-grained 3D AQI monitoring. {Some recent work in \cite{ref-22,ref-24,ref-25}, they all combine UAV and ground WSN to achieve fine-grained 3D AQI monitoring.} However, their schemes have two challenges: only small-scale regional AQI monitoring and high costs. Therefore, we need to design energy-efficient monitoring schemes.
\subsubsection{\textbf{Vision-based Monitoring Methods}} With the development of computer vision technology, vision-based methods have also been used for AQI estimation. AQNet \cite{ref-12} realizes fine-grained 3D AQI monitoring by analyzing images taken by UAV and combining them with ground sensing systems. Liu \textit{et al.} in \cite{ref-13} used a camera-enabled mobile device to crowdsource images for AQI monitoring. However, since GDPR does not allow arbitrary exchange data between agencies, crowdsourcing data collection is no longer applicable in the existing framework.


\subsection{AQI Inference Models}
Existing AQI monitoring frameworks use AQI inference models to predict real-time or future AQI distributions. These works also can generally be divided into two categories: \textit{inference by sensor data} and \textit{inference by vison data}.
\subsubsection{\textbf{Inference by sensor data}}
Shaban \textit{et al.} in \cite{ref-28} proposed an urban air pollution monitoring and forecasting system based on a support vector machine model, which uses historical air quality data collected by particle sensors to infer AQI distribution. In \cite{ref-29}, a gaussian process regression model was proposed to achieve 2D air quality monitoring. A mobile crowdsourcing based 3D probabilistic concentration estimation method (3D-PCEM) was proposed to extend the inference to 3D in \cite{ref-8}. To infer AQI distribution more accurately, many neural network-based methods have been proposed to capture the spatio-temporal correlation features in \cite{ref-30,ref-31}. However, the above work ignores the correlation between the topology of the terrestrial WSN and the distribution of AQI.
\subsubsection{\textbf{Inference by vison data}}
In \cite{ref-32}, a CNN-based model was proposed to capture the correlation between the haze image and AQI distribution. Models with the help of UAV as well as deep learning techniques are proposed for fine-grained 3D AQI monitoring in \cite{ref-26}.

In this paper, we propose two novel inference models: (1) A low-energy-consumption visual model using Dense-MobileNet that can infer image-based AQI scale in different monitoring regions; (2) A spatio-temporal inference model using GC-LSTM that can infer fine-grained AQI values in each region of the ground WSN. {Specifically, we can accurately obtain the fine-grained AQI values of different directions with the convergence of the above two models.}

\subsection{Federated Learning in UAV}
In recent years, Federated Learning (FL) models have been used to analyze private data because of its privacy-preserving features. FL is to build machine-learning models based on datasets that are distributed across multiple devices while preventing data leakage. For example, UAV swarms will employ FL models for executing various tasks such as AQI monitoring and target recognition. Zeng \textit{et al.} in \cite{ref-36} applied FL framework to design power allocation and scheduling schemes for UAV swarms. Furthermore, Google proposed a privacy-protected AQI monitoring mobile application, called VisionAir\footnote{\href{https://blog.tensorflow.org/2020/02/visionair-using-federated-learning-to-estimate-airquality-tensorflow-api-java.html}{VisionAir}}, which uses FL to estimate air quality over the whole region in a fine-grained manner. In particular, VisionAir allows users to crowdsource 3D samples without compromising privacy.

{Inspired by the previous work, we leverage the FL technique to learning models that estimate air quality by using UAV-taking photos. Specifically, FL technique not only enables agencies to learn AQI estimation models collaboratively but also enables regional-level AQI monitoring.}

\begin{figure}[!t]
	\centering
	\includegraphics[width=0.8\linewidth]{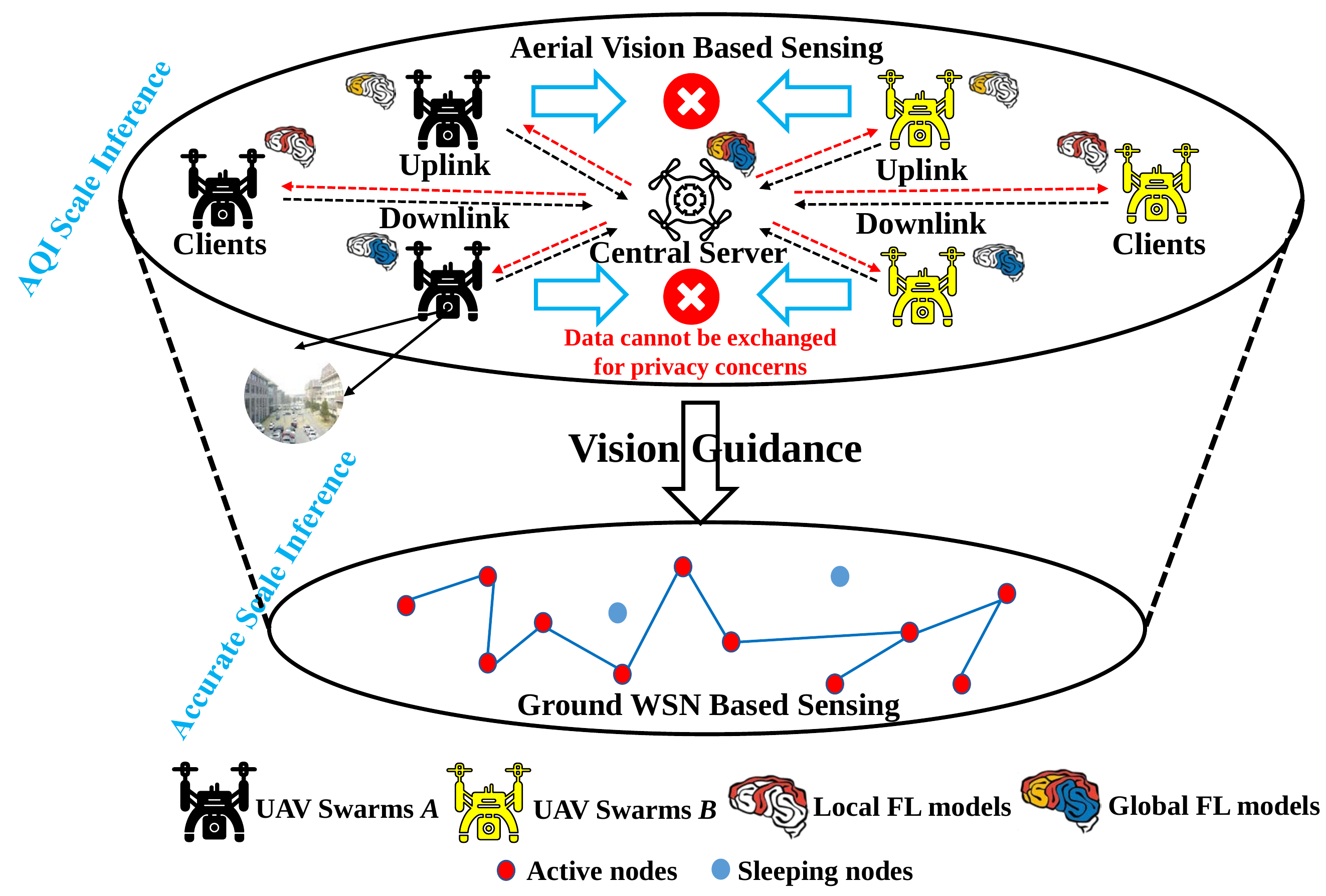}
	\caption{The overview of the light-weight federated learning-based UAV vision-guided aerial-ground air quality sensing framework.}
	\label{fig-1}
\end{figure}
\section{System Model}\label{sec-3}
As shown in Fig. \ref{fig-1}, the proposed framework includes aerial UAV swarms sensing network and ground sensing systems. Vision-based aerial UAV sensing and sensor-based ground WSN sensing form a hybrid spatio-temporal sensing network.

This framework can combine air UAV sensing and ground WSN sensing to achieve fine-grained 3D AQI monitoring. In particular, the proposed framework uses FL technique equipped with a Dense-MobileNet model to achieve region-level AQI inference without compromising privacy.

\subsection{Federated Learning Model}\label{sec-fl}
FL is a distributed machine learning (ML) paradigm that has been designed to train ML models without compromising privacy. With this scheme, UAVs can contribute to the overall model training while keeping the training data locally. Particularly, FL problem involves learning a \textit{locally} and \textit{globally} predicted model from the local dataset separately stored in dozens of or even hundreds of UAVs.

Consider a group of wirelessly connected autonomous UAVs flying at the same altitude. As shown in Fig. \ref{fig-1}, UAV swarms from \textit{A} and \textit{B} are executing AQI monitoring tasks simultaneously. The UAV swarm includes a \textit{Central Server} $\mathcal{C}$ and a set of $\mathcal{K}$ of $K$ \textit{Clients}. We assume that each UAV $k$ collects its a series of haze images (i.e., local dataset) $\mathcal{H}_k$ of size $H_k$. Therefore, we can obtain the local training dataset size $H = \sum\nolimits_{k = 1}^K {{H_k}} $. In FL model, we assume that the input sample vector with $d$ features is ${x_i} \in {\mathbb{R}^d}$ and the labeled output value for the input sample $x_i$ is ${y_i} \in \mathbb{R}$. So we can combine them into a set of input-output pairs $\{ {x_i},{y_i}\} _{i = 1}^{{H_k}}$. For AQI monitoring task, we need to learn an optimal model to infer AQI value by inputting the training sample vector $x_i$ (i.e., the haze images) and finding the model parameter vector $\omega  \in {\mathbb{R}^d}$ that characterizes the output value $y_i$ (i.e., the inference value of AQI) with loss function ${f_i}(\omega )$ (e.g., ${f_i}(\omega ) = \frac{1}{2}(x_i^T\omega  - {y_i})$). The goal is to minimize the loss function ${f_i}(\omega )$. The loss function on the data set of UAV $k$ is defined as:
\begin{equation}\label{eq-1}
{F_k}(\omega ): = \frac{1}{{{H_k}}}\sum\nolimits_{i \in {H_k}} {{f_i}(\omega )}  + \lambda h(\omega ),
\end{equation}
where $\omega  \in {\mathbb{R}^d}$ is the local model parameter, $\forall \lambda  \in [0,1]$, and $h( \cdot )$ is a regualarizer function.

To solve (\ref{eq-1}), FL framework uses a stochastic gradient descent scheme \cite{ref-34}. After UAV $k$ is minimized (\ref{eq-1}), the gradients are uploaded to the central server using uplink. At the central server, we can define the globally predicted model problem as follows:
\begin{equation}\label{eq-2}
\arg \mathop {\min }\limits_{\omega  \in {{\mathbb {R}}^d}} F(\omega ),F(\omega ) = \sum\nolimits_{k = 1}^K {\frac{{{H_k}}}{H}} {F_k}(\omega ),
\end{equation}
we recast the globally predicted model problem in (\ref{eq-2}) as follows:
\begin{equation}\label{eq-3}
\arg \mathop {\min }\limits_{\omega  \in {{\mathbb {R}}^d}} F(\omega ): = \sum\nolimits_{k = 1}^K {\frac{{\sum\nolimits_{i \in {H_k}} {{f_i}(\omega ) + \lambda h(\omega )} }}{H}} .
\end{equation}
\subsection{Aerial Sensing}
{Section \ref{sec-fl} demonstrates the FL-based aerial sensing framework. First, UAVs from different agencies use a camera to take a series of haze images in different monitoring areas. Second, a single UAV uses the Dense-MobileNet model to learn the correlation between the haze images and AQI distribution.} Each UAV uploads the gradient of a trained local model to a central server. Finally, the central server aggregates the gradients uploaded by each UAV to obtain the optimal global model. This optimal global model can realize end-to-end learning of haze images to predict region-level AQI distribution.
\subsection{Ground Sensing}
Constructing a graph structure is usually preferred to represent the spatial relationships of ground WSN. According to the definition of the graph structure, the ground sensing system uses the topological characteristics of the sensor location to construct a graph convolutional neural network (GCN). GCN combines the spatio-temporal model (i.e., LSTM) to achieve AQI inference and future air quality prediction. The ground sensing system uses the inferred results from GC-LSTM model to modify the prediction results of the aerial sensing framework to obtain real-time and future fine-grained AQI distributions.

\section{Aerial Sensing: Federated Learning-based Dense-MobileNet Model}\label{sec-4}
The aerial sensing network adopts a light-weight Dense-MobileNet model to enable UAV swarms from different agencies to perform AQI scale monitoring because (1) UAV swarms from different agencies distributed in different air positions can train AQI monitoring models collaboratively without sharing raw data, thereby greatly expanding the scope of sensing; (2) UAV can stay at different heights, collect 3D samples at different angles, and can train a FL model through wirelessly connecting and communicating; (3) UAV swarms equipped with a Dense-MobileNet model do not need to carry extra sensors, and they only need to carry a built-in camera, thereby can extend the monitoring time.

Reference \cite{ref-3} has shown that machine learning can be effectively used to estimate air quality using camera images. Although previous work used ML models to infer AQI scale, these work were learned using images taken by a few static cameras. Such approaches would result in a lack of image diversity, thereby reducing model's performance. In this work, we use a Dense-MobileNet model (i.e., a variant of CNN model) with UAVs to learn the extracted haze features to infer a fine-grained AQI scale.

In this section, we first introduce the imaging principles of haze images. Second, we detail the extraction process of haze features. Third, we introduce the details of Dense-MobileNet model. Finally, we demonstrate a fine-grained AQI inference algorithm based on light-weight federated learning.


\subsection{Overview of Haze Image Imaging Principle}\label{sec-4-1}
\textbf{\textit{Bill-Lambert Law}} can be applied to the atmosphere to describe the attenuation of sunlight and starlight as they pass through the atmosphere, as defined follows:
\begin{equation}\label{eq-4}
\beta (x) = {e^{ - \lambda d(x)}},
\end{equation}
where $x$ represents the pixel coordinates, $\beta (x)$ denotes the transmission matrix, $d(x)$ is the scene's depth map, and $\lambda $ is the extinction coefficient of the medium. Equation (\ref{eq-4}) indicates that we can estimate $\mathrm{PM}$ concentration by the extinction coefficient of the wavelength. The extinction coefficient is modeled by the following equation:
\begin{equation}\label{eq-5}
\phi (x) = {\phi _0}(x)\beta (x) + {\phi _s}(x)(1 - \beta (x)),
\end{equation}
where $\phi (x)$ respresents the pixel value sensed by UAVs with the camera, $\phi_0(x)$ denotes the brightness of the scene, and $\phi_s(x)$ respresents the airlight color vector (see below for details).

${\phi _0}(x)\beta (x)$ is used to calculate the radiation value of the light reflected from the surface of the object, which is directly transmitted to UAV camera after attenuation. Airlight in the scene can reach UAV camera after scattered by air molecules and $\mathrm{PM}$. The second term of Equation (\ref{eq-5}) ${\phi _s}(x)(1 - \beta (x)$ can be used for calculating the value of airlight. 

\subsection{Haze Feature Extraction}\label{sec-4-2}
The first step is to convert the haze images collected by the UAV camera into grayscale images, and then further convert them to binary images using the \textit{Otsu} method \cite{ref-39}. Then we extract the features with haze estimation from the binary image. To achieve end-to-end learning of the haze binary image to the AQI scale estimate by using Dense-MobileNet model, we need to find the statistical features most relevant to the haze estimation in the haze image.

In the following, we investigate six features related to haze estimation in haze binary images.

\subsubsection{\textbf{Dark Channel Prior}} 
According to Section \ref{sec-4-1}, the haze model described in the imaging principle of the haze image, we can use transmission to describe the attenuation of scene radiation. To address the attenuation problem of a single blurred vision during transmission, He \textit{et al.} in \cite{ref-42} introduced the concept of a dark channel prior. Dark channel prior assumes that for all outdoor images at least one color channel (e.g., three channels RGB), there are some pixels with zero or very low intensity. For a haze–free image $\mathcal{P}$, the dark channel prior is defined as,
\begin{equation}\label{eq-6}
{{\mathcal{ P}}^{dark}}(x) = \mathop {\min }\limits_{y \in \Omega (x)} \bigg(\mathop {\min }\limits_{c \in \{ R,G,B)} \big({{\mathcal{ P}}^c}(y)\big)\bigg),
\end{equation}             
where $\mathcal{ P}^c$ represents one of the three color channels of $\mathcal{ P}$, $\Omega (x)$ denotes all pixel colors in a local patch. Therefore, the rough approximation of the haze thickness can be estimated by the dark channel prior. According to Equation (\ref{eq-6}), the transmission is defined as follows:
\begin{equation}\label{eq-7}
\tilde \beta (x)= 1 - \mathop {\min }\limits_{y \in \Omega (x)}\big (\mathop {\min }\limits_{c \in \{ R,G,B\} } \frac{{{I^c}(y)}}{{{\phi _s}^c(x)}}\big),
\end{equation}     
where $\frac{{{I^c}(y)}}{{{\phi _s}(x)}}$ represents a haze image normalized by air light $\phi_s(x)$, and $\mathop {\min }\limits_{c \in \{ R,G,B\} } \frac{{{I^c}(y)}}{{{\phi _s}^c(x)}}$ is a dark channel prior of a normalized blurred image. 

\subsubsection{\textbf{Depth Map}}
Depth map is proposed to indicate the density of the haze. We follow reference \cite{ref-43} to defind it as follows:
\begin{equation}\label{eq-9}
\mathrm{DM}(y,x;\alpha ,\theta ) = \sum\limits_{p \in {\mathcal{ N}}} {U({y_p},x;\alpha ) + \sum\limits_{(p,q) \in \mathcal{S}} {V({y_p},{y_q},x;\theta )} }, 
\end{equation}
where $\mathcal{ N}$ represents the superpixel set, $\mathcal{S}$ represents the neighborhood superpixel pair set, $U(*)$ is the unary potential of the parameterized pixel value, $V(*)$ is the paired potential of the single layer neural network parameterization, and $\alpha$, $\theta$ are the model parameters.

\subsubsection{\textbf{Blue Channel}}
The blue of the sky can indicate how polluted the sky is. If the sky is gray, we think the sky is very polluted. HSV (Hue, Saturation, and Value) represents the properties of the color space and can be used for the extraction and segmentation of certain pixels \cite{ref-42}. Therefore, We use pixel saturation values to select hue or intensity as the main attributes to extract blue pixel features. The formal definition of converting RGB color space to HSV space is as follows:
\begin{equation}\label{eq-10}
\begin{array}{l}
\mathrm{H} = (1 + \frac{{|2R - G - B|*(2R - G - B)}}{{4[{{(R - G)}^2} + (R - B)(G - B)]}},\\
\mathrm{S} = 1 - \frac{3}{{R - G - B}}\min (R,G,B),\\
\mathrm{V} = \frac{1}{3}(R + G + B),
\end{array}
\end{equation}
{where Hue ($\mathrm{H}$) is measured by angle, with a value range of ${{0}}^\circ  \sim {{360}}^\circ $. Saturation ($\mathrm{S}$) indicates how close the color is to the spectral color. Value ($\mathrm{V}$) represents the brightness of the color, and its value range is generally 0\% (black) to 100\% (white). $\mathrm{R, G, B}$ represents the components of each color channel. We can use it to extract the blue of the haze image.}

\subsubsection{\textbf{RMS Contrast}}
Image contrast is another factor that affects the $\mathrm{PM}$ concentration of a haze image. More intuitively, humans judge the contrast of the scene through visual perception, thereby determining the quality of the weather. Since aerial lighting does not contain scene information, $\mathrm{PM}$ will cause a decrease in image contrast. We use the standard deviation of the pixel intensity of the image to describe the contrast of the image, which is defined as the root mean square (RMS) of the image,
\begin{equation}\label{eq-11}
\mathrm{RMS} = \sqrt {\frac{1}{{mn}}\sum\limits_{i = 1}^n {\sum\limits_{j = 1}^m {{{(I_{ij} - \mathrm{avg}(I))}^2}} } },
\end{equation}
in which $m$, $n$ represents the size of the image, $I_{ij}$ represents the intensity of the image at pixels $i$, $j$, and $avg (I)$ debotes the average intensity of all pixels in the image.

\subsubsection{\textbf{Image Entropy}}
The image entropy feature may provide PM-related information. It can use image texture to quantify the information contained in the image, which is defined as follows:
\begin{equation}\label{eq-12}
\mathrm{entropy} =  - \sum\limits_{i = 1}^M {{p_i}{{\log }_2}{p_i}} ,
\end{equation}
where $M$ denotes the image's maximum intensity and $p_i$ is the probability that the pixel intensity is equal to $i$. With the increase of $\mathrm{PM}$ concentration, the image becomes more and more blurred and loses details, thereby leading to a decrease in image entropy.

\subsubsection{\textbf{Image Smoothness}}
The smoothness of the sky is used to indicate how clear the sky is. We use the average of the magnitude of the gradient in the sky region to define the smoothness of the sky, as follows:
\begin{equation}\label{eq-13}
\nabla I = \frac{{\partial I}}{{\partial x}}\hat x + \frac{{\partial I}}{{\partial y}}\hat y,
\end{equation}
Where $I$ represents the intensity of the image. At the $x$ direction, $\frac{{\partial I}}{{\partial x}}$ denotes the gradient, and at $y$ direction, $\frac{{\partial I}}{{\partial y}}$ denotes the gradient. The average value of the gradient amplitude is defined as follows,
\begin{equation}\label{eq-14}
\mathrm{avg}(|\nabla I|) = \frac{1}{{mn}}\sum\limits_{i = 1}^n {\sum\limits_{j = 1}^m {\sqrt {{{(\frac{{\partial I}}{{\partial x}})}^2} + {{(\frac{{\partial I}}{{\partial y}})}^2}} } }. 
\end{equation}
where $\mathrm{avg (*)}$ is the average of the gradient amplitudes of the two-dimensional image.

\begin{figure*}[!t]
	\centering
	\includegraphics[width=0.65\linewidth]{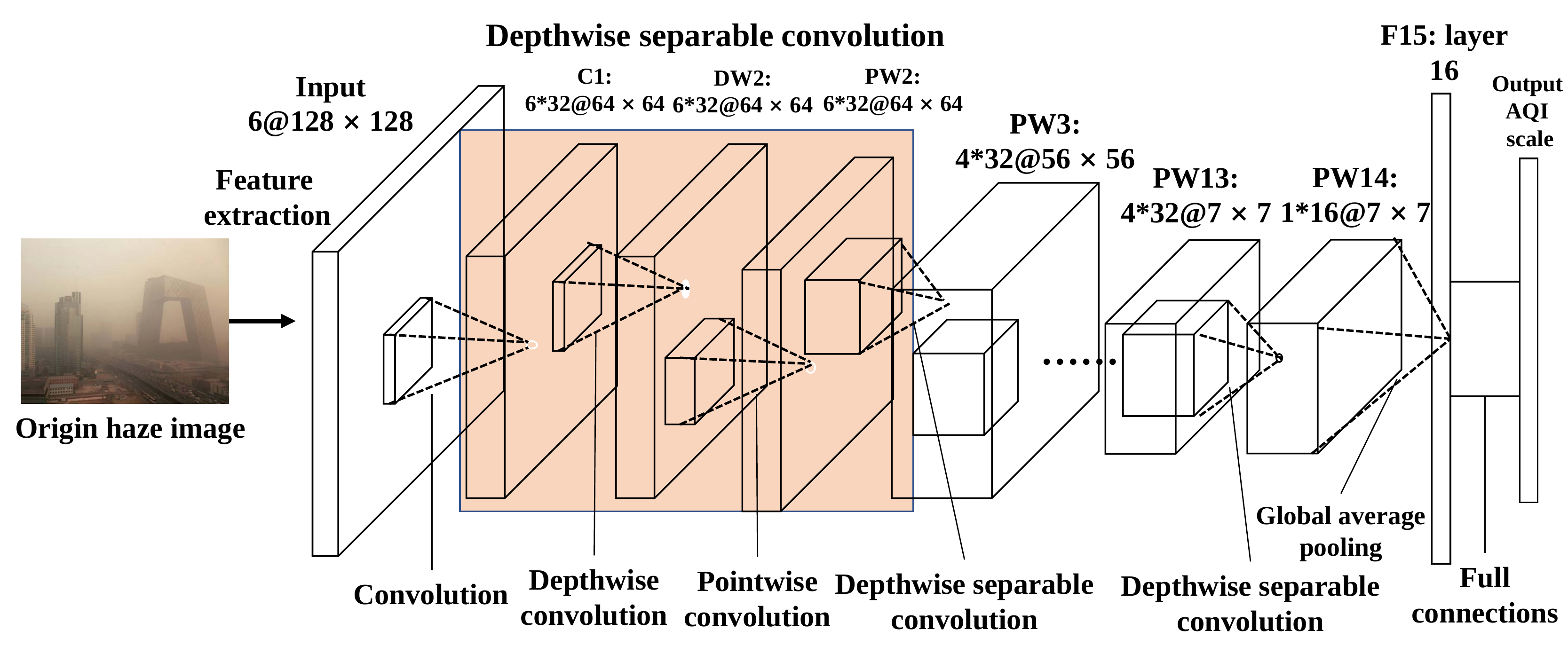}
	\caption{The architecture of the proposed Dense-MobileNet model.}
	\label{fig-3}
\end{figure*}

\subsection{Dense-MobileNet based Federated Learning for AQI Scale Inference}
In section \ref{sec-4-2}, we extracted six features related to haze estimation. Previous work has confirmed that CNN models have achieved unprecedented performance in image processing and vision applications. Furthermore, the MobileNet model is a light-weight CNN model used for mobile devices due to the reduction of the complex hierarchical structure of CNNs \cite{ref-19}. Therefore, we utilize the MobileNet model to learn these features for accurate AQI scale inference.  

To achieve the best feature learning with the fewest parameters, Dense Convolutional Network (DenseNet) \cite{ref-44} was proposed for efficient feature learning. Inspired by DenseNet, we applied the feature transfer mechanism in MobileNet to achieve efficient feature learning. Specifically, this mechanism uses the output of all the previous layers as the input of the next layer to transfer feature learning. Furthermore, for the same task, the more complex the model, the greater the energy consumption required to train the model. Therefore, to reduce the energy consumption of UAV when performing monitoring tasks, we propose Dense-MobileNet model with an efficient feature learning mechanism to achieve low energy consumption AQI scale estimation.

\textbf{Preprocessing:} First, we spatially resize each input haze image to $128 \times 128$ pixels. Second, we extracted six haze-related features and normalized them to $[0,1]$ in grayscale. To accelerate the model's convergence, pixels outside the six haze-related features are normalized to zero-mean.

\textbf{Model Architecture:} The proposed model architecture includes a feature extraction layer, several depthwise separable convolution layers, a global average pooling layer, and a fully connected layer, as shown in Fig. \ref{fig-3}. Dense-MobieNet uses deep separable convolutions to perform convolution operations. It decomposes the original convolution integral into two different convolutions, one on the feature map and the other on the channel. The feature extraction layer extracts six features related to the haze estimation in the haze image, thereby obtaining a feature tensor with a size of $128\times 128\times 6$. The extracted features are encoded as input to depthwise separable convolution layers so that the model can estimate the haze by learning the haze features. This method can improve the performance of the model by using feature coding as input instead of the entire haze image. Finally, the global average pooling layer pools the feature tensors and inputs them into the fully connected layer, thereby obtaining an estimate of the AQI scale. This model completes the end-to-end learning of haze image to AQI scale estimation.

\textbf{Light-weight Characteristics:}  We assume that the input of the convolutional layer is $M \times H \times W$ and the output is $N \times H \times W$. The size of the standard convolution kernel is $M \times K \times K \times N$ so thst the total calculation amount is $M\times H\times W\times K\times K\times N$. The calculation amounts of the Depthwise layer and the Pointwise layer in Dense-MobieNet are $M \times W \times H \times K \times K$ and $N \times W \times H \times 1 \times 1$, respectively. Therefore, the ratio of the parameter quantity and calculation quantity of the two is defined as follows:
\begin{equation}
\begin{array}{l}
\mathrm{Para} = \frac{{M \times K \times K \times 1 + M \times 1 \times 1 \times N}}{{M \times K \times K \times N}} = \frac{1}{N} + \frac{1}{{{K^2}}},\\
\mathrm{Comp} = \frac{{M \times W \times H \times K \times K \times 1 + M \times W \times H \times 1 \times 1 \times N}}{{M \times W \times H \times K \times K \times N}} = \frac{1}{N} + \frac{1}{{{K^2}}}.
\end{array}
\end{equation}
As can be seen from the ratio of the parameter amount and the calculation amount above, compared with the traditional standard convolution, the depth-dividable convolution can greatly reduce the parameter amount and the calculation amount.

\textbf{AQI Scale Inference:} According to AQI classification (i.e., $[X_{min}, X_{max}]$) system proposed by WHO , we consider AQI scale inference model as a classification problem. According to the classification index of WHO and AQI value of the collected data, we perform AQI scale pre-classification on the training data. The proposed model learns haze images and combines the inferred results of the ground perception system to achieve fine-grained AQI scale inference.

\subsection{Dense-MobileNet based Federated Learning Algorithm}
To enable public and private institutions to infer AQI scale collaboratively and expand the scope of UAV swarms, we introduce a federated learning framework to achieve this goal. We propose a federated learning-based Dense-MobileNet algorithm. It consists of four steps:
\begin{enumerate}[label=(\roman*)]
	\item {The central server model ${\omega _t}$  is initialized through pre-training that utilizes domain-specific public datasets without privacy concerns;
		\item The central server distributes the copy of the global model ${\omega _t}$ to all UAVs through downlink (i.e., $\omega _t^u \leftarrow {\omega _t}$), and each UAV trains its copy on local dataset by using local Dense-MobileNet model, i.e., ${f_u}(\omega ) = \frac{1}{{{D_u}}}\sum\nolimits_{{d_i} \in {\mathcal{D}_u}} {{f_i}(\omega )} $, where ${f_u}(\omega )$ is the local loss over datapoints at client $u$, ${\mathcal{D}_u}$ is the local dataset, ${D_u}$ is the size of local dataset, and ${f_i}(\omega )$ is the local loss function;
		\item Each UAV uploads its model updates $\omega _{t+1}^{u}$ to the central server through uplink. The entire process does not share any private data, but instead sharing the parameters;
		\item The central server aggregates the updated parameters $\omega _{t+1}^{u}$ uploaded by all UAVs to build a new global model, i.e., ${\omega _{t + 1}} = \arg \mathop {\min }\limits_{\omega  \in \mathbb{R}} \frac{1}{D}\sum\limits_{u \in \mathcal{U}} {{D_u}} {f_u}(\omega ),D = \sum\limits_{u \in \mathcal{U}} {{D_u}} $, and then distributes the new global model ${\omega _{t + 1}}$ to each UAV.}
\end{enumerate}

Let $\mathcal{U}_c$ represents the central server and $\mathcal{U}$ represents UAVs set from different institutions. The pseudocode of algorithm is presented in Algorithm \ref{al-1}:
\begin{algorithm}[t]\label{al-1}
	\caption{Federated Learning-based Dense-MobileNet Algorithm.}
	\LinesNumbered 
	\KwIn{{UAVs $\mathcal{U} = \{ {u_1},{u_2}, \cdots ,{u_k}\} $ and the central server $\mathcal{U}_c$. The mini-batch size $m$, the number of iterations $n$ and the learning rate $\alpha$.}}
	\KwOut{Parameter $\omega$.}
	{$\mathcal{U}_c$ initializes the global model $\omega_{t}$ by training public dataset}\;
	\ForEach{round $i = 1,2,3,\cdots, n$}
	{
		{$\{u\} \leftarrow $ select UAVs $\mathcal{U}$ from different institutions to participate in this round of training}\; 
		\While{$\omega_t$ has not convergence}
		{
			\ForEach{UAV $u_k \in {u}$ \textbf{in parallel}}
			{
				{Initalize $\omega _{t+1}^{u}  \leftarrow  \omega _{t}^u$\;
					Conduct a mini-batch input haze images $\{ x^{(i)}\} _{i = 1}^m$\;
					Conduct a mini-batch labels $\{ {y}^{(i)}\} _{i = 1}^m$\;
					Features are extracted according to Eq. (\ref{eq-6})--(\ref{eq-14})\;
					${f_u}(\omega ) = \frac{1}{{{D_u}}}\sum\nolimits_{{d_i} \in {\mathcal{D}_u}} {{f_i}(\omega )} $\;
					$\omega _{t + 1}^u \leftarrow \omega _t^u - \alpha {f_u}(\omega )$\;}
			}
		}
		{	$\mathcal{U}_c$ collects the all parameters from $\{u\}$ to update ${\omega _{t + 1}}$\;
			${\omega _{t + 1}} = \frac{1}{D}\sum\limits_{u \in \mathcal{U}} {{D_u}} {f_u}(\omega )$\;}
	}
	\Return $\omega $
\end{algorithm}

\begin{figure*}[!t]
	\centering
	\includegraphics[width=0.65\linewidth]{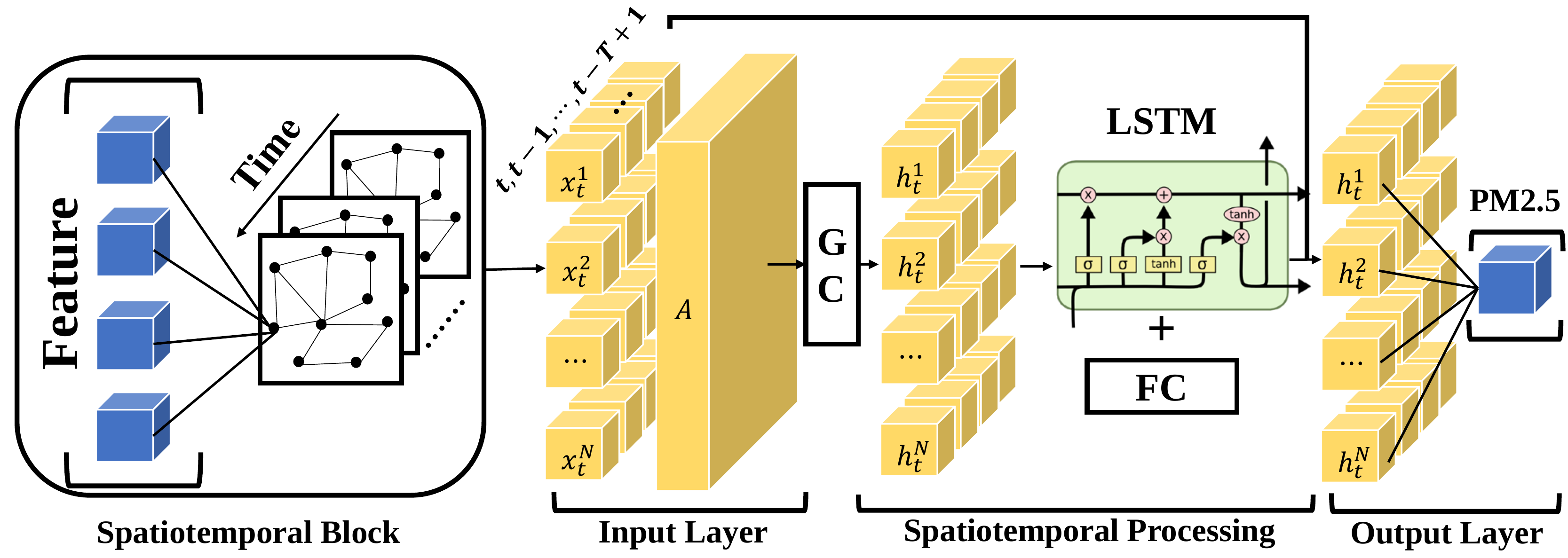}
	\caption{The architecture of the proposed GC-LSTM model.}
	\label{fig-4}
\end{figure*}
\section{Ground Sensing: AQI Inference By Garph-based Groud Sensor Monitoring}\label{sec-5}
We use ground-sensing WSN to achieve accurate realtime and future AQI  inference from the perspective of spatio-temporal. We observe the following facts: (1) There is a topology relationship between the positions of different ground-sensing sensors; (2) There is a temporal relationship between historical data on the same static station, and a spatial relationship between different static stations; (3) Historical observation data at different static stations can be defined as graph signals. Inspired by the above facts, we will construct a graph neural network to achieve fine-grained AQI spatio-temporal prediction. Furthermore, ground-sensing networks provide accurate prior knowledge for aerial sensing inference.

In this work, we propose a graph convolutional neural network-based long short-term memory (GC-LSTM) model to achieve accurate real-time and future AQI spatio-temporal inference, as illustrated in Fig. \ref{fig-4}. This model architecture includes a spatio-temporal block, an input layer, a spatio-temporal processing layer (i.e., an LSTM block), and an output layer.

\subsection{Spatial Dependency Modelling}
\subsubsection{Graph Construction} Each monitoring station can be regarded as a node in the graph, and the relationship between the monitoring stations can be represented by the edges in the graph. According to the definition of the graph structure, the input features of the $N$ monitoring stations at the time step $t$ can be translated to graph signals. We assume that $M$ is the number of features associated with each node, and the feature matrix can be expressed as ${X_t} \in {\mathbb{R}^{N \times M}}$. We use an undirected graph $G = (V, E, \mathcal{A})$ to represent the spatial relationship between stations in different locations, where $N$ nodes ${v_i} \in V$ and each edge $({v_i},{v_j}) \in E$. Let $\mathcal{A} \in {\mathbb{R}^{N \times N}}$ denote the spatial weight matrix and ${\mathcal{A}_{i,j}}$ denote the spatial correlation between $v_i$ and $v_j$.

We construct $\mathcal{A}$ from the spatial distance between AQI monitoring stations. The formal definition is as follows:
\begin{equation}\label{eq-15}
{\mathcal{A}_{i,j}} = \left\{ \begin{array}{l}
\frac{1}{{{d_{i,j}}}},i \ne j\\
0,\mathrm{otherwise}
\end{array} \right.
\end{equation}
where $d_{i,j}$ is the distance between position $i$ and position $j$ calculated from latitude and longitude.
\subsubsection{Spectral Graph Convolutional Networks} The core idea of the graph convolutional neural network (GCN) is to use the Laplacian spectral matrix and the convolution kernel in CNN to define the spectral convolution kernel. Therefore, the propagation function between convolutional layers in GCN is defined as follows:
\begin{equation}\label{eq-16}
{J^{(l + 1)}} = \sigma ({\tilde D^{ - \frac{1}{2}}}{ \mathcal{\tilde A}}{\tilde D^{ - \frac{1}{2}}}{J^{(l)}}{\omega ^{(l)}}) 
\end{equation}
{where $ \mathcal{\tilde A} = \mathcal{A} + \mathcal{L}$ ($\mathcal{L}$ is the identity matrix), ${\tilde D}$ is the degree matrix of $ \mathcal{\tilde A}$ (i.e., ${{\tilde D}_{ii}} = \sum {j{{ \mathcal{\tilde A}}_{ij}}} $), $J$ is the feature of each convolutional layer, $\omega^{l}$ is the weight martic, and $\sigma$ is a non-linear activation function. Note that ${{\tilde D}^{ - \frac{1}{2}}} \mathcal{\tilde A}{{\tilde D}^{ - \frac{1}{2}}}$ is similar to a symmetric normalized Laplace matrix.}

\subsection{Temporal Dependency Modelling}
Since the historical air pollution data observed by ground monitoring stations usually exist in the form of time series, we can use time series prediction models to achieve AQI inference. Previous work generally used support vector machine models to predict AQI. However, these models cannot capture the spatio-temporal dependencies within the parameters. Therefore, we introduce a sequence model in GCN to dynamically capture the spatio-temporal dependence of parameters and historical data. 

In this work, we use a variant of a recurrent neural network, called LSTM, to achieve accurate AQI inference, as shown in Fig. \ref{fig-4}. LSTM uses a well-designed ``gate'' structure to remove or add information to the state of the cell. The ``gate'' structure is a method of selectively passing information. LSTM cells include forget gates $f_t$, input gates $i_t$, and output gates $o_t$. The calculations on the three gate structures are defined as follows:
\begin{equation}\label{eq-17}
\begin{array}{l}
{f_t} = {\sigma _l}({W_f} \cdot [{h_{t - 1}},{x_t}] + {b_f}),\\
{i_t} = {\sigma _l}({W_i} \cdot [{h_{t - 1}},{x_t}] + {b_i}),\\
{{\tilde C}_t} = \tanh ({W_C} \cdot [{h_{t - 1}},{x_t}] + {b_C}),\\
{C_t} = {f_t}*{C_{t - 1}} + {i_t}*{{\tilde C}_t},\\
{o_t} = {\sigma _l}({W_o} \cdot [{h_{t - 1}},{x_t}] + {b_o}),\\
{h_t} = {o_t}*\tanh ({C_t}).
\end{array}
\end{equation}
where ${W_f},W_i,W_C, W_o$, and ${b_f},{b_i},{b_C},{b_o}$ are the weight matrices and the bias vectors for input vector $x_t$ at time step $t$, respectively. $\sigma_l$ is the activation function, $*$ represents element-wise multiplication of a matrix, $C_t$ represents the cell state, $h_{t-1}$ is the state of the hidden layer at time step $t-1$, and $h_t$ is the state of the hidden layer at time step $t$.

\subsection{Spatiotemporal Prediction}
Given the topological map structure of ground monitoring stations and the spatio-temporal correlation of historical data, we propose a GC-LSTM model to predict real-time and future AQI values in different regions:
\begin{equation}\label{eq-18}
[AQ{I_{(t + 1)}}, \cdots ,AQ{I_{(t + T')}}] = h([{x_{t - T + 1}}, \cdots ,{x_t}];G(V,E,\mathcal{A})).
\end{equation}

As shown in Fig. \ref{fig-4}, we tranlate four features: AQI historical observations, meteorological variables, temporal predictors, and spatial predictors into graph signals. We use graph convolution (GC) operations and LSTM blocks to extract spatial and temporal features, respectively. In the proposed model, we use the graph convolution feature of the graph signal in series as the input of LSTM block.

For each spatio-temporal block, we extract the graphic signal $x_t$ and the spatial weight matrix $\mathcal{A}$ at each time step $t$ to calculate the spatial feature $h_t$ through the graph convolution kernel. Next, the graphic signal $x_t$ and $h_t$ are connected as the input of LSTM block. Finally, the output of LSTM block is used as the input of the fully connected (FC) layer, and the output of FC layer is the real-time or future AQI value.

\begin{figure}[t]
	\centering
	\large
	\subfigure []{\includegraphics[width=0.4\linewidth]{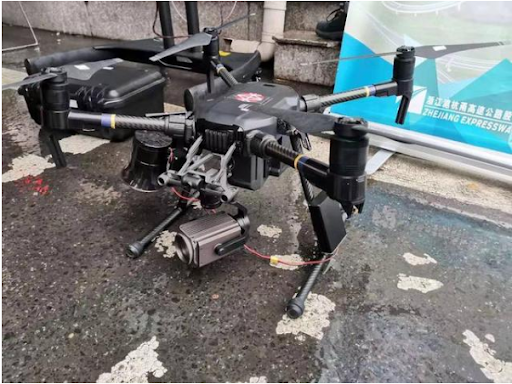}
		\label{a-1}}
	\hfill
	\subfigure[]{	\includegraphics[width=0.4\linewidth]{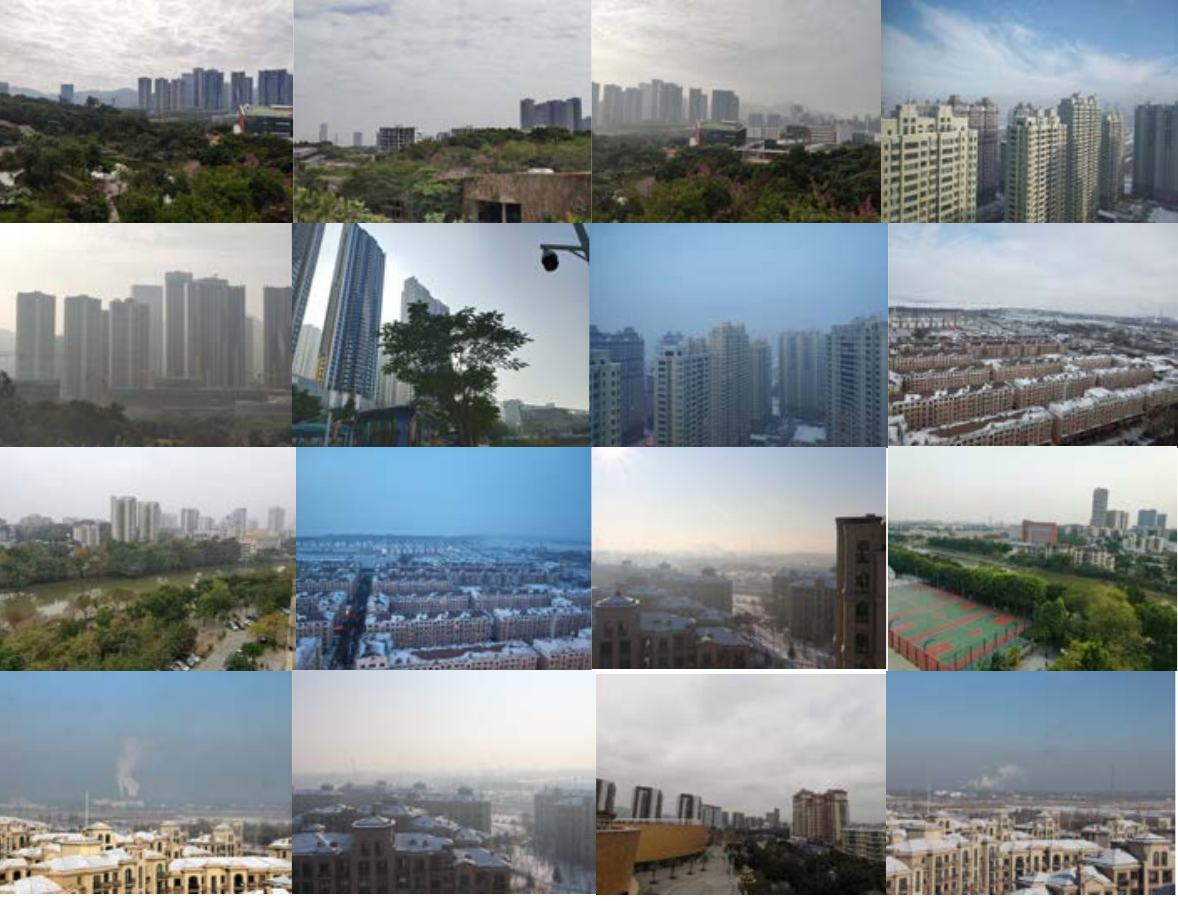}
		\label{b-1}}
	\caption{(a) The UAV for vision-based sensing. (b) An overview of the aerial image dataset: more than 5,000 labeled haze images taken in Guangzhou, Shenzhen, Hangzhou, Harbin, and Daqing.}
	\label{fig-5}
\end{figure}
\section{Experiments}\label{sec-6}
In this section, we evaluate the performance of the proposed framework on a real-world dataset. Specifically, we first introduce the equipment components in the proposed framework, including ground devices and aerial devices. We then provide the details about how to collect aerial haze image data.

\textbf{Ground Devices:} To facilitate modeling, we use static monitoring stations deployed in urban areas (e.g., Hangzhou, Harbin, Guangzhou, and Shenzhen) as ground sensing devices. We use the longitude and latitude of the monitoring stations as their location and treat locations as nodes in the graph structure \footnote{Longitude and latitude data are available at \href{http://www.weather.com.cn/air/}{http://www.weather.com.cn/air/}.}. Furthermore, the longitude and latitude can also determine the distance between stations, thereby deriving the weight matrix $\mathcal{A}$. 

\textbf{Aerial Devices:} As shown in Fig. \ref{a-1}, we choose $\mathrm{MAVIC}$ 2 $\mathrm{ZOOM}$ UAVs from DJI company as the aerial sensing devices. The maximum journey distance  of these UAVs is  18 Km each time, and the maximum remote control distance is 8 Km. These UAVs can utilize GPS sensors to provide real-time 3D monitoring of positions and use 4K cameras to collect haze image data. The $\mathrm{MAVIC}$ 2 $\mathrm{ZOOM}$ UAV's maximum battery life is about 30 minutes, and it can hover in the air to collect data at a ${270^ \circ }$ viewing angle. The UAVs can communicate with each other through wireless networks, and the maximal communication distance of UAVs is 7$\sim$9 Km.

\subsection{Evaluation Setup and Data Description}
According to the definition of graph structure, we use the ground monitoring stations deployed by the government to construct the graph neural network. Nine UAVs were used to collect haze image data, of which one was the central server and the others were divided into two UAV swarms. 

\textbf{Ground Sensing Data:} The historical observations of AQI from the monitoring stations can be downloaded from \href{http://www.weather.com.cn/air/}{http://www.weather.com.cn/air/}. In this paper, historical AQI observations during the first six months of 2019 are used to evaluate the framework. We select the air quality data from the first five months as the training dataset and the data from the sixth month as the test dataset. Furthermore, since the air quality data is time-series data, we need to use them at the previous time interval, i.e., ${x_{t - 1}},{x_{t - 2}}, \cdots ,{x_{t - r}}$, to predict the air quality at time interval $t$, where $r$ is the length of the history data window.

\textbf{Aerial Sensing Data:} We respectively collected 5,298 haze images in Hangzhou, Guangzhou, Shenzhen, Harbin, and Daqing including two-level distribution covering ``good'' to ``dangerous'', as illustrated in Fig. \ref{b-1}. These images can be used for AQI estimation. We randomly divide the image dataset into a training and a test set at a ratio 8:2.

\textbf{Index of Performance:} We adopt Root Mean Square Error (RMSE) to indicate the robustness of inference as follows:
\begin{equation}\label{eq-19}
\mathrm{RMSE} = {[\frac{1}{n}\sum\limits_{i = 1}^n {{{(|{y_i} - {{\hat y}_p}|)}^2}} ]^{\frac{1}{2}}}.
\end{equation}
where $y_i$ is the observed air quality, and ${\hat y}_p$ is the predicted air quality.

\begin{figure}[t]
	\centering
	\large
	\subfigure []{\includegraphics[width=0.45\linewidth]{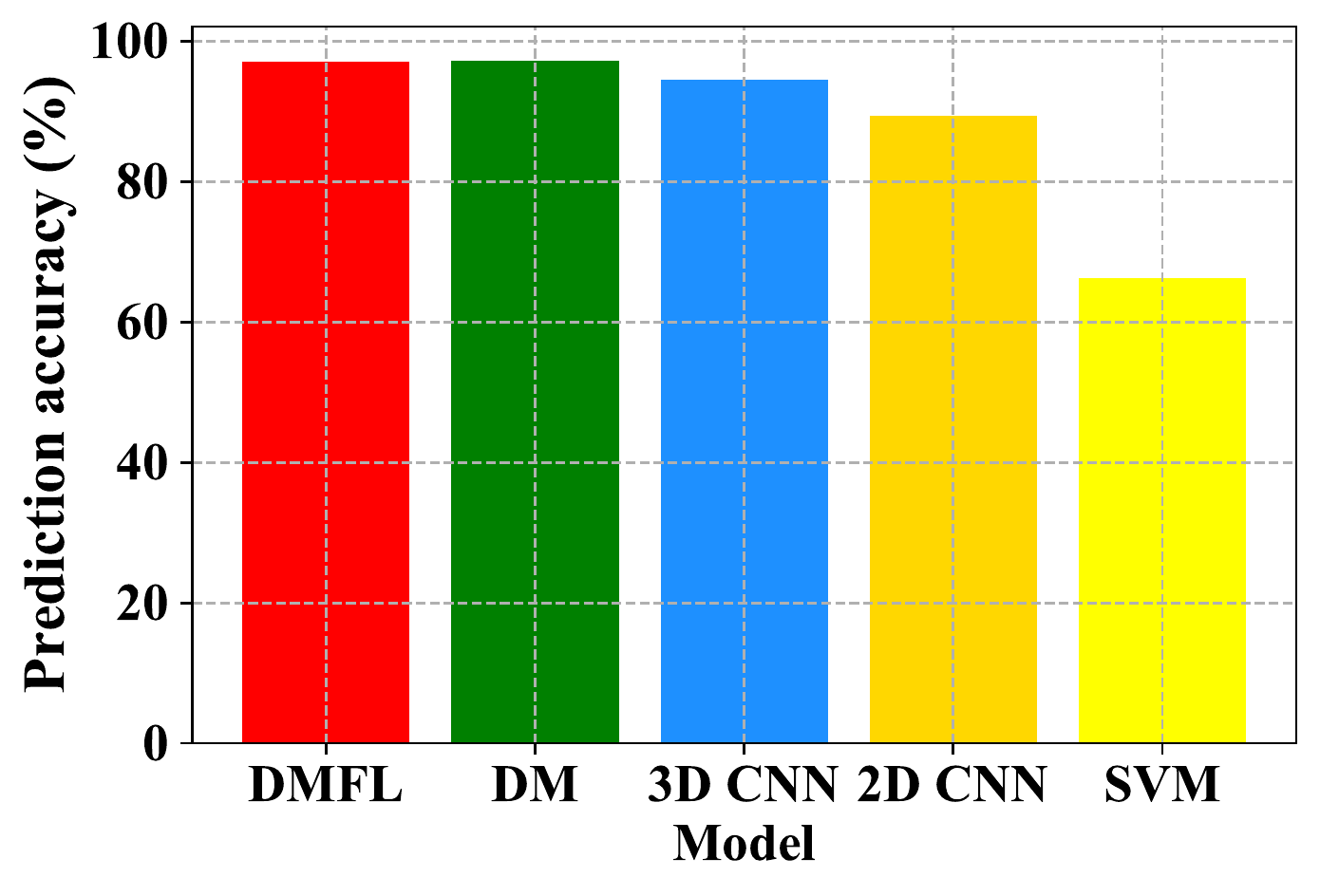}
		\label{a-11}}
	\hfill
	\subfigure[]{	\includegraphics[width=0.45\linewidth]{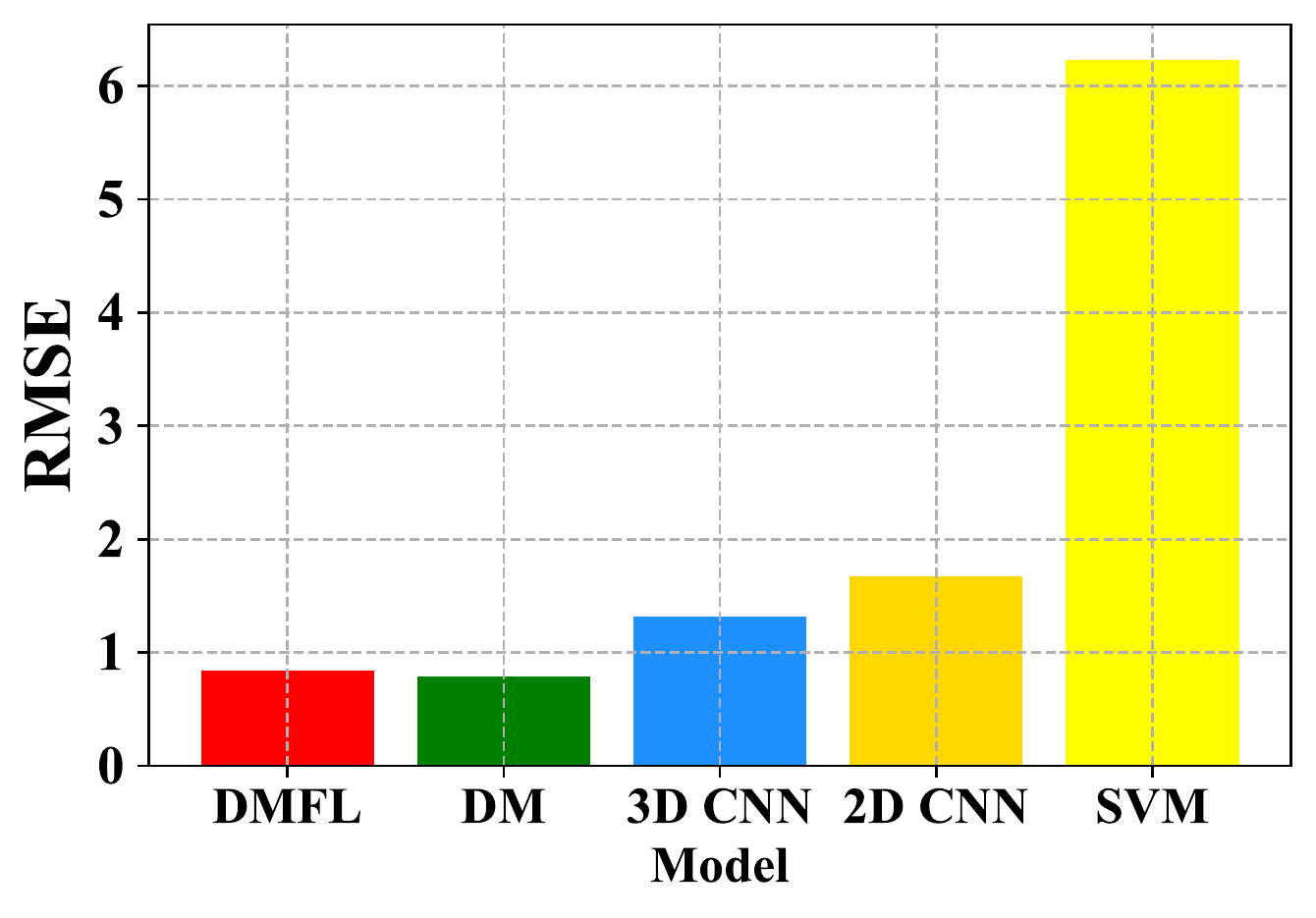}
		\label{b-11}}
	\caption{Performance comparison between the proposed model and other models: (a) Prediction accuracy; (b) RMSE.}
	\label{fig-6}
\end{figure}
\subsection{AQI Inference Accuracy}
\subsubsection{Vision-based Aerial Sensing}
We compare the performance of the proposed ``Dense-MobileNet + Federated Learning'' (DMFL) model with those models of DM, 3D CNN, 2D CNN, and Support Vector Machine (SVM) under an identical simulation configuration. Among these five competing methods, DMFL is a federated machine learning model, and the rest are centralized ones. The models of 3D CNN and 2D CNN are widely-adopted image-based models that {have} good performance for AQI scale inference tasks, and SVM is a popular machine learning model for general prediction applications.

In all investigations, we use the same haze image dataset. The prediction results are given in Fig. \ref{a-11} for real-time AQI scale prediction. We can observe that the accuracy of DMFL is higher than those of 3D CNN, 2D CNN, and SVM but lower than that of DM. Specifically, the accuracy of DMFL is 97.13\% lower than that of the  Dense-MobileNet (DM) in this experiment. This result is contributed by the fact that DMFL inherits the advantages of DM's outstanding performance in prediction tasks. This means that the proposed DMFL can achieve high-precision predictions without compromising privacy.

Fig. \ref{b-11} shows the inference robustness comparison between DMFL and other models. We can find that $\mathrm{RMSE}$ results of DMFL model are very close to that of DM {model} and much better than those of other models.
The reason is that the core technique of DMFL for AQI prediction is the DM structure, so the performance of DMFL is comparable to DM model. Furthermore, the DMFL can protect data privacy by keeping the training dataset locally. The above results verify that the proposed model not only achieves high-precision inference without compromising privacy but also maintains better robustness.

\begin{table}[!t]
	\centering
	\caption{Performance Comparsion of RMSE For GC-LSTM, GRU, LSTM, and AQNet}
	\begin{tabular}{ccccc} 
		\toprule
		Metrics  &Real-time	&After 2 hours	&After 4 hours	&After 6 hours		\\
		\midrule
		\textbf{GC-LSTM} &\textbf{3.212} &\textbf{4.589}  &\textbf{6.357}   &\textbf{9.145}   \\
		GRU      &4.025	&9.156	&14.892   &22.047	\\
		LSTM      &4.217	&9.635	&15.068   &21.624	\\
		AQNet       &4.493	&9.264	&17.695    &24.753		\\
		\bottomrule
	\end{tabular}
	\label{tb-1}
\end{table}

\subsubsection{Sensor-based Ground Sensing} 
To further improve the fine-graininess of aerial sensing results, we build a GC-LSTM model on the ground to capture the historical observation and the spatio-temporal dependency of future AQI. Therefore, we compare GC-LSTM with state-of-the-art sequence models (i.e., Gate Recurrent Unit (GRU) \cite{ref-45}, LSTM Net \cite{ref-47}, AQNet \cite{ref-12}) to illustrate the performance of GC-LSTM. {GRU and LSTM are variants of Recurrent Neural Network (RNN). They are generally used in sequence prediction and natural language processing tasks. AQNet is a 3D fine-grained spatiotemporal sequence prediction model, which includes ground sensing system and air sensing system.} In this paper, we use the GC-LSTM model and other models to predict AQI values in real-time and in the future (i.e., after 2, 4, 6 hours, respectively).

The results about $\mathrm{RMSE}$ are presented in Table \ref{tb-1}, {which} shows that GC-LSTM's error is much smaller than other models. This is because GC-LSTM captures not only the spatial correlation of historical data on monitoring stations but also the temporal dependence between air quality data. Moreover, other models have ignored the topological correlation between the monitoring stations, while we use this topological correlation to build a graph convolutional neural network in our work.

\begin{figure}[t]
	\centering
	\large
	\subfigure []{\includegraphics[width=0.45\linewidth]{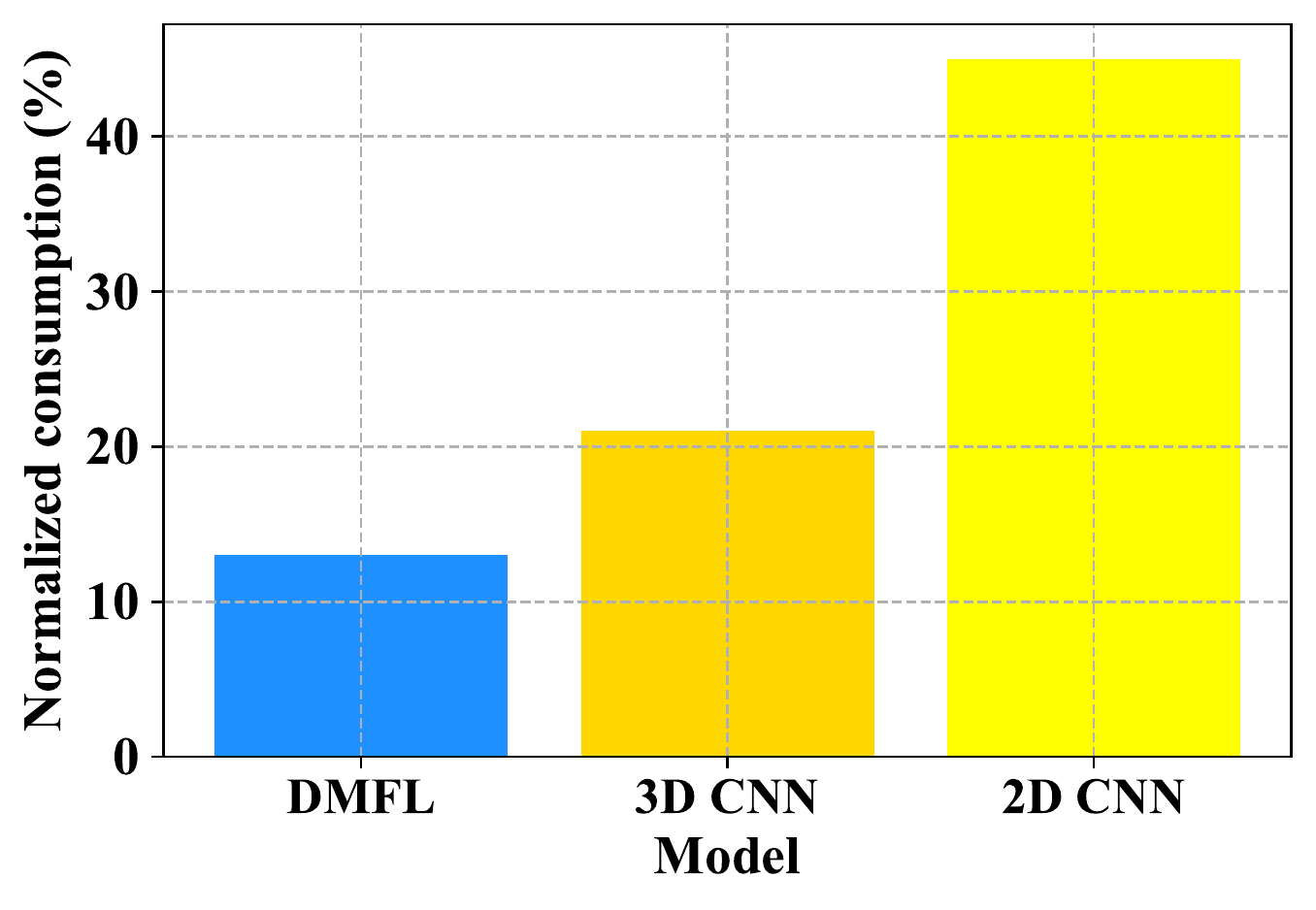}
		\label{a}}
	\hfill
	\subfigure[]{	\includegraphics[width=0.45\linewidth]{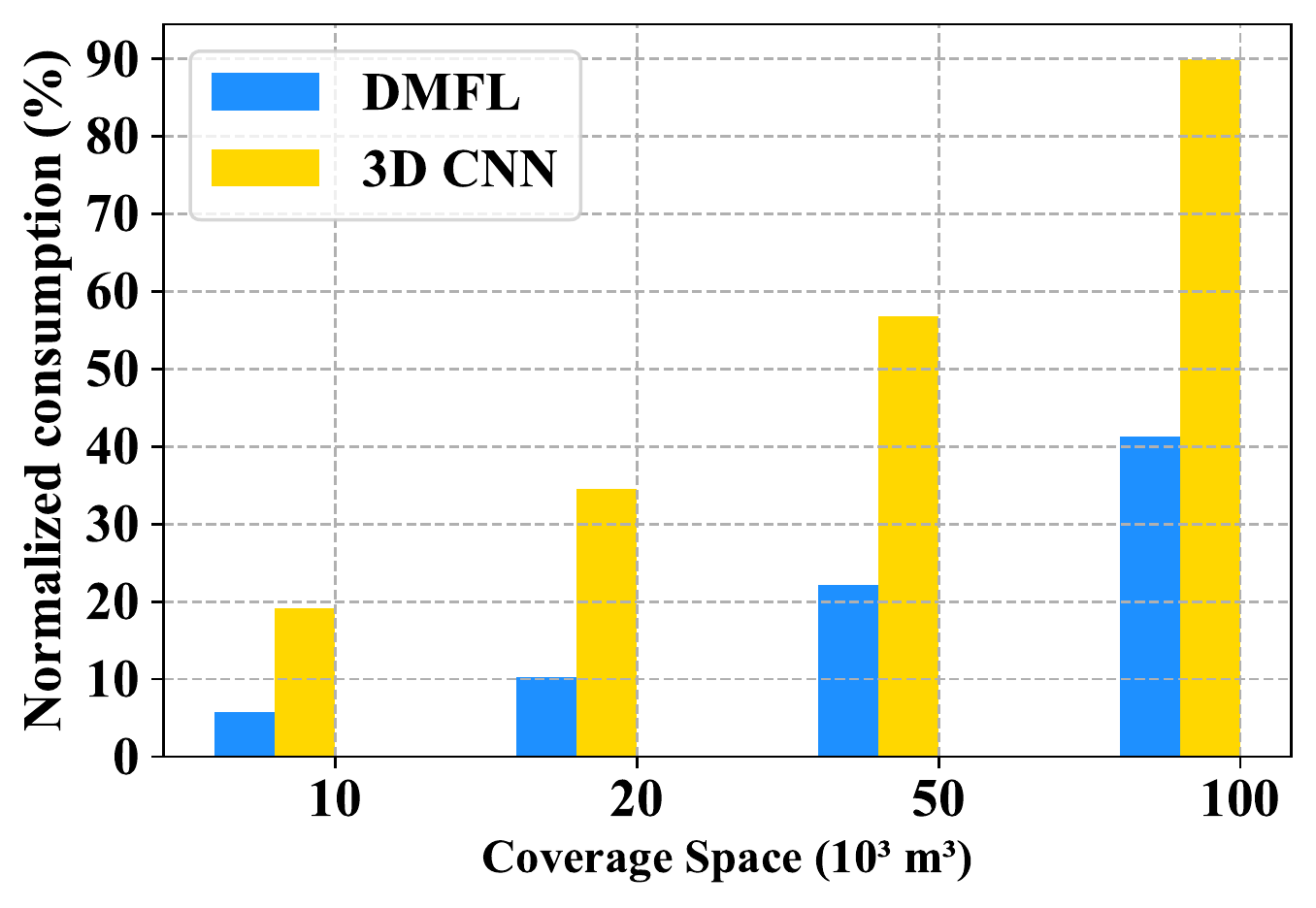}
		\label{b}}
	\caption{The energy-efficiency comparison between the proposed model and other models: (a) Training task; (b) Monitoring task.}
	\label{fig-7}
\end{figure}

\subsection{Energy Efficiency}
{Due to the limited battery life of UAVs, we need to design a low-energy monitoring scheme for UAVs to complete more tasks in the air. In this paper, we use a federated learning-based DM model to reduce UAV's energy consumption in the air.} We analyze the energy consumption of UAVs from two aspects: (1) the energy consumed by model training; (2) the energy consumed under the same monitoring scope. We compare the energy consumption of the models by using a UAV equipped DMFL model and other models (i.e., 3D CNN and 2D CNN models) to perform the same task, respectively. We use normalized consumption to indicate the level of energy efficiency.

\textbf{Training Task:} We assume that UAV uses different models to complete {the} model training in the air. As shown in Fig. \ref{a}, DMFL model requires the least energy to complete a model training. This reason is that the DMFL reduces the hierarchical structure of CNN and also uses the Dense structure to share the features of each layer. Both 3D CNN and 2D CNN achieve the high-precision AQI scale inference by adding the hierarchical structure of CNN.

\textbf{Monitoring Task:} Fig. \ref{b} shows the energy consumption of different models under the same monitoring scope. It shows that when the monitoring range is 100 $m^3$, the energy consumption of DMFL is about half that of 2D CNN. This is because the federal learning framework allows UAV swarms to collaborate on learning. The above results illustrate the superior performance of the proposed model with low energy consumption.

{\subsection{Discussion}
	In this section, we discuss the performance of the proposed framework from the perspectives of privacy, accuracy, and energy consumption as follows:}
{
	\begin{itemize}
		\item \textbf{\textit{Privacy:}} In the proposed framework, institutions or organizations do not need to access or collect user data and they do not need to share data with their partners, which protects user privacy. Indeed, different organizations only need to share model updated parameters instead of sharing user data.
		\item \textbf{\textit{Accuracy:}} From the experimental results, the proposed framework achieves the state-of-the-art results in both aerial sensing and ground sensing. Ground sensing can provide a more accurate numerical reference for aerial sensing. The aerial sensing system uses a light-weight Dense-MobileNet model, whose performance exceeds that of large-scale models such as the 3D CNN model.
		\item \textbf{\textit{Energy Consumption:}} It is very important for UAV swams to achieve low energy consumption. In the framework we designed, we use an improved lightweight CNN model suitable, i.e., MobileNet model for mobile phones to achieve fine-grained 3D AQI monitoring. Such a design not only realizes high-precision AQI monitoring, but also reduces UAV energy consumption. This provides a research direction for the future deployment of large-scale UAV swarms for AQI monitoring.
\end{itemize}}
\section{Conclusion}\label{sec-7}
In this paper, we propose an aerial-ground air quality sensing framework with UAV swarms for monitoring and forecasting the air quality in a fine-grained manner with privacy-preservation federated learning. First, we use the light-weight Dense-MobileNet model to learn the haze feature of UAV-taking haze images to achieve energy-efficient AQI scale inference. Second, to further improve the inference accuracy of the aerial sensing system, we propose a GC-LSTM model based on graph topology to realize real-time and future AQI {predictions.} We evaluate the performance of the proposed framework on a real-world dataset and compare it with 3D CNN, 2D CNN, and SVM methods that may compromise privacy during forecasting. The results show that the proposed method is better than the existing methods. {To the best of our knowledge, this is one of the pioneering works for air quality forecasts with federated deep learning.}

{In the future, we will focus on the communication efficient technique and novel model compression technique of the FL framework \cite{ref-48}. First, the large number of communication rounds between UAVs results in expensive communication overhead. Due to the limitation of UAV power, expensive communication overhead makes UAV unable to achieve long-term AQI monitoring. Therefore, we need to find a communication efficient FL framework to solve this challenge. Second, in general, the size of deep learning models is very large and difficult to apply to micro mobile devices, such as UAV. In order to be able to deploy large-scale and complex deep learning models to UAVs, we need to design some novel model compression techniques to achieve this goal.}


%




\ifCLASSOPTIONcaptionsoff
\newpage
\fi



%
\bibliographystyle{IEEEtran}
\bibliography{reference}

%




\begin{IEEEbiographynophoto}{Yi Liu}
	(S'19) received the B.Eng degree from the Heilongjiang University, China in 2019. He is now pursuing his Ph.D. degree in Monash University, Australia. His research interests mainly focus on security and privacy protection in machine learning, wireless communications, and networking.
\end{IEEEbiographynophoto}
\begin{IEEEbiographynophoto}{Jiangtian Nie}
	(S'18) received her B.Eng degree with honors in Electronics and Information Engineering from Huazhong University of Science and Technology, Wuhan, China, in 2016. She is currently working towards the Ph.D. degree with ERI@N in the Interdisciplinary Graduate School, Nanyang Technological University, Singapore. Her research interests include incentive mechanism design in crowdsensing and game theory. 
\end{IEEEbiographynophoto}
\begin{IEEEbiographynophoto}{Xuandi Li}
	received the B.S. degree from Beijing Institute of Technology, Zhuhai in 2013. She is currently a research engineer in Nanyang Technological University, Singapore. Her research interests mainly focus on blockchain and Internet of Things.
\end{IEEEbiographynophoto}
\begin{IEEEbiographynophoto}{Syed Hassan Ahmed}
	(SM'18) is currently working at JMA Wireless as a Product Specialist for Distributed Antenna System (DAS), CBRS, Small Cell, and virtualized RAN product line. Previously, he was an Assistant Professor in the Department of Computer Science at Georgia Southern University, USA. 
\end{IEEEbiographynophoto}
\begin{IEEEbiographynophoto}{Wei Yang Bryan Lim}
	graduated with double First Class Honours in Economics and Business Administration (Finance) from the National University of Singapore (NUS) in 2018. He is currently an Alibaba PhD candidate with the Alibaba-NTU Joint Research Institute, Nanyang Technological University, Singapore. His research interests include Federated Learning and Edge Intelligence.
\end{IEEEbiographynophoto}
\begin{IEEEbiographynophoto}{Chunyan Miao}
	received the B.S. degree from Shandong University, Jinan, China, in 1988, and the M.S. and Ph.D. degree in Nanyang Technological University, Singapore, in 1998 and 2003, respectively. She is currently a Professor in the School of Computer Science and Engineering at Nanyang Technological University (NTU), and the Director of the Joint NTUUBC Research Centre of Excellence in Active Living for the Elderly (LILY). Her research focus on infusing intelligent agents into interactive new media (virtual, mixed, mobile and pervasive media) to create novel experiences and dimensions in game design, interactive narrative and other real world agent systems.
\end{IEEEbiographynophoto}




\end{document}